\documentstyle[epsfig,bm,amsmath,amssymb,aps,12pt,tighten]{revtex}

\allowdisplaybreaks[3]

\def\Fs#1{{}\kern-.45em \not \kern-.12em #1\hspace{.2pt}}
\def\as{\alpha_s}
\def\A{{\cal A}}
\def\B{{\cal B}}
\def\G{{\cal G}}

\begin{document}

\tighten

\title{
\hfill\parbox{3cm}{\normalsize\raggedleft DESY 01-xxx\\ hep-ph/01xxxxx}\\[30pt]
The Process \boldmath $\gamma_L^*+ q \to (q\bar{q}g)+q$\,:\\
Real Corrections to the Virtual Photon Impact Factor}
\author{J.~Bartels\thanks{supported by the EU
  TMR-Network `QCD and the Deep Structure of Elementary Particles',
  contract number FMRX-CT98-0194 (DG 12-MIHT).}, 
S.~Gieseke$^{*}$\thanks{supported by the Graduiertenkolleg
  `Zuk\"unftige Entwicklungen der Teilchenphysik'.}, 
A.~Kyrieleis$^{\dagger}$\\[10pt]}
\bigskip
\address{II.~Institut f\"ur Theoretische Physik, Universit\"at
  Hamburg\\
  Luruper Chaussee 149, 22761 Hamburg, Germany}
\maketitle

\bigskip
\begin{abstract}%
  We calculate, for the longitudinally polarized virtual photon, the
  cross section of the process $\gamma^*+q\to (q\bar{q}g)+q$ at high
  energies with a large rapidity gap between the fragmentation system
  $q\bar{q}g$ and the other quark.  This process provides the real
  corrections of the virtual photon impact factor in the next-to
  leading order. Evidence is given for the appearance of a new $q\bar
  qg$ Fock-component of the photon state.
\end{abstract}

\section{Introduction}
\begin{figure}[t]
  \begin{center}
    \epsfig{file=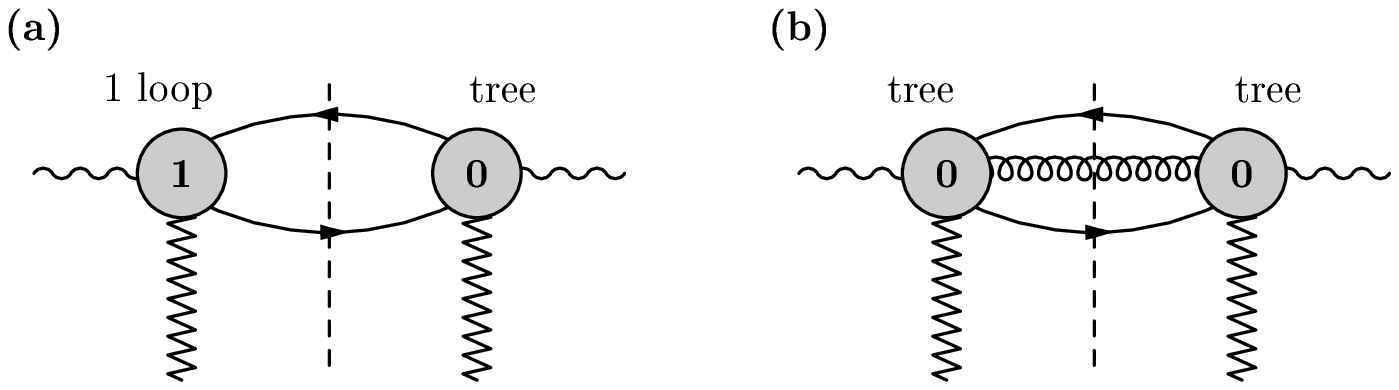,width=13cm}
    \caption{Different contributions to the $\gamma^*$ impact factor
      at NLO. \label{fig:contribif}}
  \end{center}
\end{figure}

\noindent
After the completion of the NLO virtual corrections \cite{BGQ}, the
next step in our program of computing the virtual photon impact factor
in next-to-leading order is the calculation of the emission of a real
gluon, as sketched in Fig.~\ref{fig:contribif}. We will compute these
corrections from the high energy limit of the process $\gamma^*+q \to
(q\bar qg)+q$ (Fig.~\ref{fig:qqgkin}) at leading order in $\as$. We
are interested in the region of phase space, were the $q\bar q$-pair
and the $g$ are emitted in the fragmentation region of the virtual
photon with a large rapidity gap relative to the outgoing lower quark.
We impose no further restrictions on the mass of the $(q\bar
qg)$-state, and therefore our results will also include the production
of the gluon in the central region: this corresponds to the
leading-order BFKL calculation, where the gluon is emitted with a
large rapidity gap relative to both the lower quark and to the upper
$q\bar q$-pair. For the calculation of the photon impact factor, this
piece has to be subtracted from our results.

A study of the process $\gamma^*+q \to (q\bar qg)+q$ includes the
subprocess $\gamma^*+g \to q\bar qg$ with an off-shell incoming gluon.
Together with the NLO corrections to the production process $\gamma^*
+ g \to q \bar q$ which has been calculated in \cite {BGQ}, it
provides the complete NLO corrections to the photon-gluon fusion
process with off-shell incoming bosons. This generalizes the usual
on-shell calculation of the collinear factorization scheme to the
$k_T$-factorization scheme \cite{kt-Factorization}. During the last
years evidence has been collected that, for certain final states in
the small-$x$ region, the use of this scheme provides a better
description of HERA data than the traditional collinear factorization
scheme: for the modeling of final states it is therefore important to
have the off-shell photon-gluon fusion not only in LO but also in NLO.

Finally, it is well-known that the leading order photon impact factor
can be interpreted in terms of the photon wave function and a color
dipole cross section \cite{Mue,NZC49}.  In recent years this color
dipole picture has proven to be extremely useful for analyzing HERA
data in the small-$x$ region, in particular the low-$Q^2$ region of
the $F_2$ structure function and the diffractive final states
\cite{GBW1,GBW2,GBW3}. The validity of the color dipole picture (or,
maybe, its generalization to higher multipole terms) beyond leading
order therefore presents an issue which has to be investigated. In
NLO, the validity of the photon wave function picture is one of the
results that should come out of the NLO photon impact factor
calculation: in the interpretation of our results we will therefore
pay particular attention to this aspect.

\section{Kinematics and calculational techniques}
\label{sec:qqgkin}

\begin{figure}[t]
  \begin{center}
    \epsfig{file=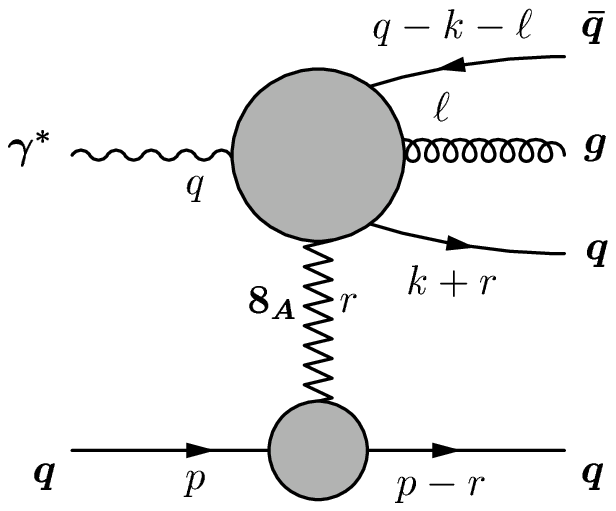,width=5cm}
    \caption{Kinematics of the process $\gamma^* + q \to q\bar{q}g + q$.
      \label{fig:qqgkin}}
  \end{center}
\end{figure}

\noindent
The notations of the scattering process are depicted in
Fig~\ref{fig:qqgkin}. The incoming momenta $q = q' - xp$ and $p$ are
used to construct the Sudakov basis $q',p$ (with ${q'}^2=p^2=0$) for
the other momenta, and $s= 2 p\cdot q'$ denotes the center of mass
energy of the incoming particles.  The virtuality of the photon is
$q^2 = -Q^2$, and as usual, $x=Q^2/s$. The momentum transfer is
denoted by $r$ (with $r^2=t$), the outgoing gluon carries momentum
$\ell$ and the momenta of the quark and antiquark are $k+r$ and
$q-k-\ell$, respectively.  We express these momenta in terms of their
Sudakov variables,
\begin{align}
  \label{eq:sudakovqqg}
  k &= \alpha q' + \beta p + k_\perp\;,\\
  \ell &= \alpha_\ell q' + \beta_\ell p + \ell_\perp\;,\\
  r &= \alpha_r q' + \beta_r p + r_\perp\;, 
\end{align}
where the Euclidean notation is used for the two dimensional
transverse momenta, e.g.\ $\bm k^2 = -k_\perp^2$. We are interested
in the limit of large energy $s$: $s$ has to be much larger than
$Q^2$, $\bm k^2$, $\bm \ell^2$, and $\bm r^2$, and the $\alpha$ variables of
the $q\bar{q}g$-system are of the order unity. The regions of
integration are
\begin{align} 
{\cal O}(t/s)< \alpha< 1 \nonumber \\ 
\frac{|\bm \ell|}{\sqrt{s}} < \alpha_{\ell} <1 - \alpha.
\end{align}

For the produced gluon we require a large rapidity gap between the gluon 
and the lower quark. In other words, we include both the
fragmentation region of the virtual photon 
\begin{equation}
\alpha_{\ell \; 0} < \alpha_{\ell} < 1-\alpha
\end{equation}
(where $\alpha_{\ell \; 0}$ denotes a small but energy independent 
cutoff parameter to be specified later),
and the `upper' part of the central region
\begin{equation}
\frac{|\bm \ell|}{\sqrt{s}} < \alpha_{\ell} < \alpha_{\ell\; 0}. 
\end{equation}
There is a remaining part of the phase space which we do not consider: 
the `lower' half of the central region:
\begin{equation}
\frac{\bm \ell^2}{s} < \alpha_{\ell} <  \frac{|\bm \ell|}{\sqrt{s}},
\end{equation}
and the fragmentation region of the lower quark. The central region belongs to 
the LO BFKL calculation, whereas the fragmentation region of the 
lower quark is part of the quark impact factor \cite{Fadin-qif}. 
The definition of the `upper' and `lower' parts of the central region is somewhat 
arbitrary. We have chosen the `symmetric' separation:  
$\alpha_{\ell}= \frac{|\bm \ell|}{\sqrt{s}}$ corresponds to the symmetric point   
$M^2_u = (q+r)^2=M_d^2 = (p+\ell-r)^2= \bm \ell^2 s$.
 
From the on-shell conditions of the outgoing quarks and gluons we obtain 
$\alpha_r=t/s$ and
\begin{align}
  \label{eq:betas}
  \beta_\ell s&= \frac{\bm \ell^2}{\alpha_\ell}\;,\\
  \beta s&= - \frac{(\bm k +\bm \ell)^2}{1-\alpha-\alpha_\ell}
    - \frac{\bm \ell^2}{\alpha_\ell} - Q^2\;,\\
  \beta_r s&= \frac{(\bm k+\bm r)^2}{\alpha} 
    + \frac{(\bm k +\bm \ell)^2}{1-\alpha-\alpha_\ell}
    +\frac{\bm \ell^2}{\alpha_\ell} + Q^2.
\end{align}
The Feynman diagrams are listed in Fig.~\ref{fig:realgraphs}.  The
graphs in the second line (denoted by $B_1,\ldots, B_5$) are obtained
from the diagrams of the first line ($A_1,\ldots, A_5$) by
interchanging the quark and antiquark, i.e. by transforming $k+r \to
q-k-\ell$. In terms of the Sudakov components this transformation
reads:
\begin{figure}[t]
  \begin{center}
    \epsfig{file=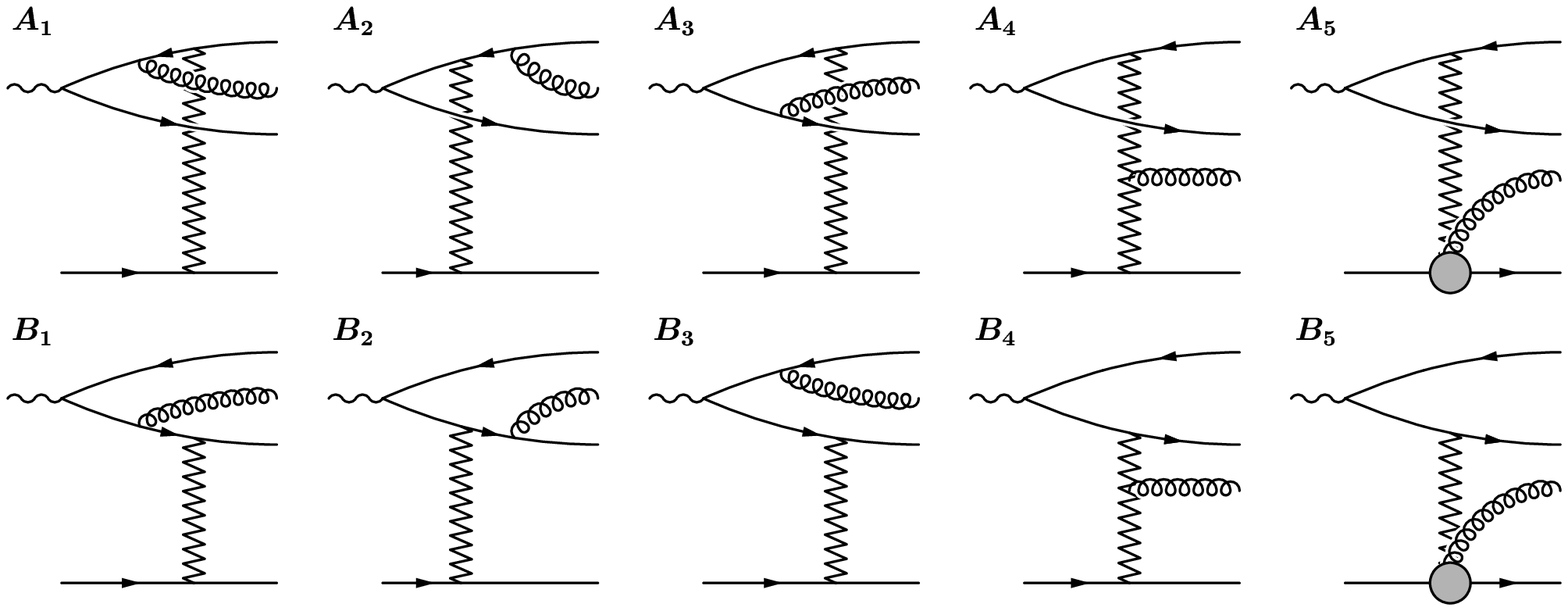,width=16cm}
  \end{center}
  \caption{Feynman diagrams for the process $\gamma^* + q \to q\bar{q}g +
    q$.}
  \label{fig:realgraphs}
\end{figure}
\begin{align}
  \label{eq:barr1}
  \alpha & \to 1-\alpha-\alpha_\ell\\
  \label{eq:barr2}
  \bm k & \to -(\bm k + \bm r + \bm \ell). 
\end{align}
In correspondence with this transformation we introduce the shorthand 
notation:   
\begin{align}
  \label{eq:bardefvars}
  \alpha_1 &\equiv \alpha\;,\\
  \bar \alpha_1 &\equiv (1-\alpha-\alpha_\ell)\;,\\
  \alpha_2 &\equiv (1-\alpha)\;,\\
  \bar \alpha_2 &\equiv (\alpha+\alpha_\ell).
\end{align}
where the bars stand for the transformation \eqref{eq:barr1},
\eqref{eq:barr2}. The denominators, that will occur during the calculation, 
have the form:
\begin{align}
D_1&=(k+\ell+r-q)^2 \nonumber\\
&={}-\bar{\alpha}_1 Q^2 - (\bm k + \bm\ell + \bm r)^2 -
\frac{\bar{\alpha}_1}{\alpha_1}\, (\bm k + \bm r)^2 -
\frac{\bar{\alpha}_1}{\alpha_\ell} \, \bm\ell^2\label{eq:D1}\\
D_2&=(k+r-q)^2 \nonumber\\
&={}-\alpha_2 \, Q^2 - \frac{1}{\alpha_1} \,
(\bm k + \bm r)^2\label{eq:D2}\\
D_3 &=(k+\ell+r)^2\nonumber\\
&={}-(\bm k + \bm\ell + \bm r)^2+ 
\frac{\bar{\alpha}_2}{\alpha_1} \,
(\bm k + \bm r)^2 + \frac{\bar{\alpha}_2}{\alpha_\ell}\, \bm\ell^2 
\nonumber\\
&={}\frac{1}{\alpha_\ell \, \alpha_1}\, (-\alpha_\ell \,
(\bm k + \bm r) + \alpha_1\, \bm\ell )^2\\
D_4&=(k-q)^2\nonumber\\
&={}-\bm{k}^2 + \frac{\alpha_2}{\bar{\alpha}_1} \, 
(\bm k + \bm\ell)^2
+ \frac{\alpha_2}{\alpha_\ell} \, \bm\ell^2 \nonumber \\
&={}\frac{1}{\alpha_\ell \, \bar{\alpha}_1} \,
(\alpha_\ell \, (\bm k + \bm\ell) + \bar{\alpha}_1 \, \bm\ell)^2\\
D_5&=(\ell-r)^2\nonumber\\
&={}-\alpha_\ell \, Q^2 - \frac{\alpha_\ell}{\alpha_1} \,
(\bm k + \bm r)^2 - \frac{\alpha_\ell}{\bar{\alpha}_1} \, 
(\bm k + \bm\ell)^2 -
(\bm\ell-\bm r)^2\\
D_6&=k^2\nonumber\\
&={}-\alpha_1 \, Q^2 - \bm{k}^2 - \frac{\alpha_1}{\bar{\alpha}_1}
\, (\bm k + \bm\ell)^2  - \frac{\alpha_1}{\alpha_\ell} \, \bm\ell^2\\
D_7&=(k+\ell)^2\nonumber\\
&={}-\bar{\alpha}_2 \, Q^2 - \frac{1}{\bar{\alpha}_1} \,
(\bm k + \bm\ell)^2\label{eq:D7}\;.
\end{align}
Under the transformation \eqref{eq:barr1}--\eqref{eq:barr2} the
denominators $D_1$ to $D_7$ obey the following relations:
\begin{align}
  \label{eq:dentranfo}
  D_1 &= \bar D_6\;,\\
  D_2 &= \bar D_7\;,\\
  D_3 &= \bar D_4\;,\\
  D_5 &= \bar D_5\;.
\end{align}

Rather than beginning with the scattering amplitude we immediately
turn to the squared matrix elements, averaged (summed) over the
helicities and color of incoming (outgoing) quarks. The polarization
vectors of the photon are
\begin{align}
  \label{eq:gammapols}
  \varepsilon_L &= \frac{1}{Q}(q'+xp) \;,\\
  \varepsilon_\pm &= \frac{1}{\sqrt{2}}(0, 1, \pm i, 0). 
\end{align}
In fact, since $\varepsilon_L = q/Q + 2xp/Q$, we make use of gauge 
invariance and simply put $\varepsilon_L = 2xp/Q$. In this paper we restrict 
ourselves to the longitudinal polarization; the transverse case will be 
presented in a subsequent paper.  

Our calculations are performed in the Feynman gauge. As to the internal 
($t$-channel) gluon numerators, we use the decomposition 
\begin{equation}
  \label{eq:gmunu}
  g_{\mu\nu} = \frac{2}{s}\left(p_\mu q'_\nu + p_\nu q'_\mu\right) +
  g_{\mu\nu}^\perp. 
\end{equation}
In the high energy limit we retain only the first term, since the 
contributions of the other terms will be suppressed by powers of $t/s$.  
For the
diagrams where the produced gluon couples to the lower fermion line, this 
may not be immediately obvious, since some of the light cone components of the 
$t$ channel gluon are not small; we have checked the
approximation against a calculation involving the full numerator, and 
we  have obtained the same answer.  
In the diagrams $A_5$ and $B_5$ we have combined two
Feynman graphs (Fig.~\ref{fig:effvertex}) into one effective diagram. Since
the produced gluon is far from the lower quark in rapidity, we can 
simplify the emission vertex:
\begin{figure}[t]
  \begin{center}
    \epsfig{file=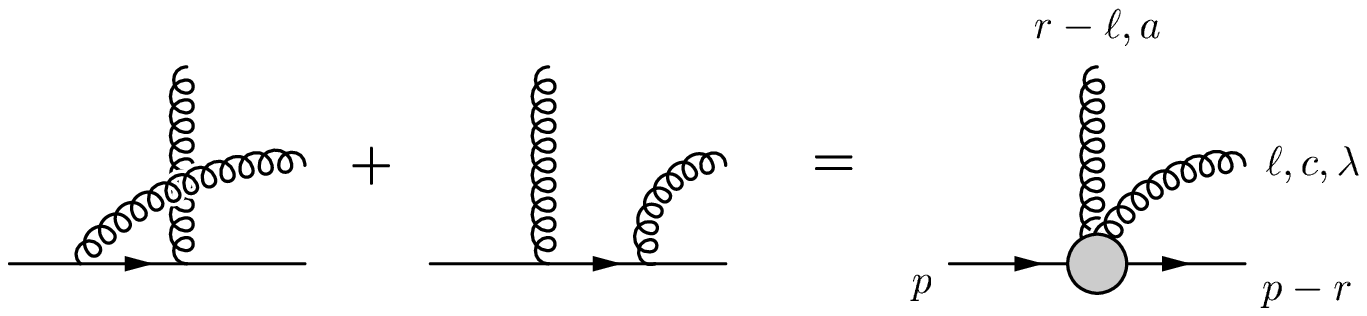,width=11cm}
  \end{center}
  \caption{Emission of a gluon from the lower quark line.  
    \label{fig:effvertex}}
\end{figure}
\begin{equation}
  \label{eq:approx5}
  \bar u(p-r, h') \left[
    \frac{\Fs q'(\Fs p-\Fs \ell)\gamma^\lambda}{(p-\ell)^2} t^a t^c
  + \frac{\gamma^\lambda (\Fs p+\Fs \ell -\Fs r)\Fs q'}{(p+\ell-r)^2}
    t^c t^a
    \right] u(p, h) 
    \approx \frac{2 p^\lambda}{\alpha_\ell}[t^c, t^a]   \delta_{hh'}\;.
\end{equation}
Here we have used that $\ell$ has large components only in the $q'$
direction; $h$ and $h'$ denote the helicities of the lower incoming
and outgoing quarks, respectively, and the Lorentz index $\lambda$ is
to be contracted with the polarization vector of the outgoing gluon
with color index $c$. $t^a$ and $t^c$ are the color group generators
in the fundamental representation, and $a$ denotes the color index
carried by the $t$-channel gluon.

The result for the helicity-summed square of the matrix element (averaged 
over helicity and color of the incoming quark) has the 
following form:
\begin{equation}
\label{eq:m2}
|{\cal M}|^2 = e^2e_f^2g^4\,\delta^{ab}\;
16 \, Q^2 \, s^2 \sum_{i,j=1}^{5} \; \left( \A\A_{ij} \,+\,\A\B_{ij} 
    \,+\, \B\A_{ij} \,+\,  \B\B_{ij}\right)\;\frac{1}{\bm{r}^4}\;\frac{g^2\delta^{ab}}{2N_c}, 
\end{equation}
where $e e_f$ is the charge of quark flavour $f$, and $g$ is the
strong coupling constant.  Following the labeling of
Fig.~\ref{fig:realgraphs} we denote the product of graph $A_i$ with
graph $A_j$ by $\A\A_{ij}$ etc.  Furthermore, the phase space 
over the final state (including the $\delta$-functions from the
on-shell conditions) is given by:
\begin{equation}
\label{eq:ps3}
d\Phi_{4} = \frac{d\alpha \, d \alpha_{\ell}\,d^{D-2} \bm k\, d^{D-2} \bm \ell\, d^{D-2} \bm r}{8s (2\pi)^{3D-4} 
\alpha_1 \bar \alpha_1 \alpha_{\ell}} \;.
\end{equation} 
$D$ denotes the space time dimension.  We calculate our results in
dimensional regularization with transverse dimension $D-2=2-2\epsilon$. 

The sum on the rhs of \eqref{eq:m2}, together with the phase space
\eqref{eq:ps3} and the flux factor $1/(2s)$ represents the
(helicity-summed) cross section of the process $\gamma^*+q \to (q \bar
q g)+q$: 
\begin{equation}
  \label{eq:crosssecdef}
  d\sigma = \frac{1}{2s} |{\cal M}|^2 d\Phi_{4}\;.
\end{equation}
The NLO real corrections to the photon impact factor will be obtained
from the sum on the rhs of \eqref{eq:m2} by doing the integral over
the $q\bar qg$ phase space and invariant mass. However, we will have
to subtract the production of the gluon in the central region
(multiregge kinematics).  This piece belongs to the LO BFKL
calculation and must not be counted twice (see below).  Integrating
over the momentum of the outgoing gluon leads to infrared
singularities which cancel against the infrared singularities of the
virtual corrections \cite{FM}. Since the calculation of the virtual
corrections has been done using the dimensional regularization, we
will use these same techniques also for the real corrections.

In order to obtain the NLO jet cross section for the off-shell
photon-gluon fusion process $\gamma^* + g \to (2\,\mbox{jets},
3\,\mbox{jets}$), we have to combine the sum in \eqref{eq:m2} with the
virtual NLO corrections to $\gamma^* + g \to q \bar q$, using a
suitable jet definition.

%%%%

\section{Results}
\label{sec:results}

\noindent
In this section we list our results for the helicity-summed matrix element.
The traces have been calculated with the help of the Mathematica
package FeynCalc \cite{feyncalc}. The results have then been simplified further by 
casting them into a factorizing form: in order to have the photon wave 
function interpretation we have tried to find a representation in which 
each term $\A\A_{ij}$ can be written as a sum of products of two terms: 
one factor belongs to the 
incoming photon, the other one to the outgoing photon, and in between we have 
a piece which belongs to the cross section which describes the interaction of 
the $q \bar q g$-system with the lower quark. 
After some algebra we arrive at the following results:  
\begin{align}
\A\A_{11}=&{}\;
   C_F \;\frac{2}{D_1^2\,D_2^2}\,(1-\epsilon)\,\alpha_1\,\bar{ \alpha }_1\,\bm{A}_1^2
      \label{eq:AA11}\\
\A\A_{12}=&{}\;
   - \frac{1}{N_c} \;\frac{1}{D_1\,D_2^2\,D_4}\left(\,(1-\epsilon)\,\alpha_1\,
        \alpha_2 \,  \bm{A}_1 \,\bm{A}_2  - 
     \alpha_1\,\alpha_2^2\,\bar{ \alpha }_1\,\bm{r}^2 \right) \\
\A\A_{13}=&{}\;
  C_F\;\frac{2}{D_1^2 \,D_2\,D_3}\,\left( (1-\epsilon)\,
        \bar{ \alpha }_1^2 \, \bm{A}_1\,\bm{A}_3\, + \bar{ \alpha }_1^2\,\left( \bm{k} + \bm{r} \right) \,
        \left( \bm{k} + \bm{\ell} + \bm{r} \right)  \right)  \\
\A\A_{14}=&{}\;
   - N_c \; \frac{1} {D_1\,D_2^2\,D_5}\,\bigg( \frac{1}{2\,
          \alpha_\ell}\,
            \alpha_1\,\left( \alpha_2\,\bar{ \alpha }_1 + 2\,(1-\epsilon)\,\alpha_\ell^2 
      \right)  \bm{A}_1\,\bm{A}_4
   \nonumber \\* 
   &+ \frac{1}{2}\alpha_1\,\alpha_2\,
          \bar{ \alpha }_1^2\,D_2 + 
       \frac{1}{2\,\alpha_\ell}\,\alpha_1\,\alpha_2^2\,
          \bar{ \alpha }_1\, \bm{A}_1 \bm{r}- 
       \alpha_1\,\alpha_2^2\,\bar{ \alpha }_1\,\bm{r}^2 \bigg)
     \\
\A\A_{15}=&{}\;
   N_c\;\frac{1}{\alpha_\ell\,D_1\,D_2^2\,D_5}\; \alpha_1\,\alpha_2^2\,\bar{ \alpha }_1^2\,
    \bm{r}^2 \\
\A\A_{22}=&{}\;
   C_F\; \frac{2}{\bar{ \alpha }_1\,D_2^2\,D_4^2}\;(1-\epsilon)\,\alpha_1\,\alpha_2^2\,\bm{A}_2^2 \\
\A\A_{23}=&{}\;
   - \frac{1}{N_c} \;\frac{1} {D_1\,D_2\,D_3\,D_4}\left( (1-\epsilon)\,\alpha_2\,
        \bar{ \alpha }_1 \,  \bm{A}_2\,\bm{A}_3\, + 
     \alpha_2\,\bar{ \alpha }_1\,
      \left( \bm{k} + \bm{r} - \alpha_1\, \bm{r} \right) \,
      \left( \bm{k} + \bm{\ell} + \bar{ \alpha }_1\, \bm{r} \right) \right)\\
\A\A_{24}=&{}\;
  N_c\;\frac{1}{D_2^2\,D_4\,D_5}\bigg(
  \frac{1}{2\,\alpha_\ell\,\bar{ \alpha }_1} \, 
          \alpha_1\,\alpha_2\,
          \left( \alpha_2\,\bar{ \alpha }_1 + 2\,(1-\epsilon)\,\alpha_\ell^2
            \right)  \bm{A}_2\,\bm{A}_4\, \nonumber \\*
     &+ \frac{1}{2} \alpha_1\,\alpha_2^2\,\bar{ \alpha }_1\,
          D_2   + 
       \frac{1}{2\,\alpha_\ell}\,\alpha_1\,\alpha_2^2\,
          \bar{ \alpha }_1\, \bm{A}_2 \bm{r} + 
       \alpha_1\,\alpha_2^2\,\bar{ \alpha }_1\,\bm{r}^2 \bigg) 
      \\
\A\A_{25}=&{}\;
   - N_c\; \frac{1}{\alpha_\ell\,D_2^2\,D_4\,D_5}\; \alpha_1\,\alpha_2^3\,\bar{ \alpha }_1\,\bm{r}^2 \\
\A\A_{33}=&{}\;
  C_F \;\frac{2}{\alpha_1\,D_1^2\,D_3^2}\;(1-\epsilon)\,\bar{ \alpha
  }_1^3\,\bm{A}_3^2 \\
\A\A_{34}=&{}\;
  N_c \;\frac{1}{D_1\,D_2\,D_3\,D_5}\,\bigg( \frac{1}{2\,\alpha_\ell}
          \bar{ \alpha }_1\,\left( \alpha_1\,\bar{ \alpha
  }_1-\alpha_\ell - 2\,(1-\epsilon)\,\alpha_\ell^2 \right)\,\bm{A}_3\,\bm{A}_4\, 
   \nonumber \\* 
    &- \frac{1}{2}\, \alpha_1\,\bar{ \alpha }_1^2\,\bar{ \alpha }_2\,
          D_2 + 
       \alpha_1\,\alpha_2\,\bar{ \alpha }_1^2\,
        D_7  + 
       \frac{1}{2\,\alpha_\ell}\,\alpha_1\,\alpha_2\,
          \bar{ \alpha }_1^2\, \bm{A}_3 \bm{r}+
  \alpha_1\,\alpha_2\,\bar{ \alpha }_1^2\,\bm{r}^2 \nonumber \\* 
  &+ \frac{3}{2}
  \,\alpha_2\, \bar{ \alpha }_1\, \bm{A}_3 \left( \bm{k} + \bm{\ell} \right)
\bigg) \\
\A\A_{35}=&\,
   - N_c\; \frac{1}{\alpha_\ell\,D_1\,D_2\,D_3\,D_5}\,\alpha_1\,\alpha_2\,\bar{ \alpha }_1^2\,
       \bar{ \alpha }_2\,\bm{r}^2 \\
\A\A_{44}=&{}\;
   N_c \;\frac{1}{D_2^2\,D_5^2}\, \left( \frac{2}{\bar{ \alpha }_1}\,\bm{A}_4^2\,\alpha_1\,
            \left( (1-\epsilon)\,\alpha_\ell^2 + \alpha_2\,\bar{ \alpha }_1 \right)
               + 
         \alpha_1\,\alpha_2\,\alpha_\ell\,\bar{ \alpha }_1\,
          D_2 + 2\,\alpha_1\,\alpha_2^2\,
          \bar{ \alpha }_1\,\bm{r}^2 \right)   \\
\A\A_{45}=&{}\;
   - N_c \;\frac{1}{D_2^2\,D_5^2} \alpha_1\,\alpha_2^2\,\bar{ \alpha }_1\,
       \bm{r}^2 \\
\A\A_{55}=&\;0 
\end{align}
\begin{align}
\A\B_{11}=&{}\;
 \frac{1}{N_c}\; \frac{1}{D_1\,D_2\,D_6\,D_7} \left((1-\epsilon)\,\alpha_1\,\bar{
  \alpha }_1 \,  \bm{A}_1\,\bm{B}_1 + 
     \alpha_1\,\bar{ \alpha }_1\,
      \left( \bm{k} + \alpha_1\,\bm{r} \right) \,
      \left( \bm{k} + \bm{\ell} + \bar{ \alpha }_2\,\bm{r} \right) \right) \\
\A\B_{12}=&{}\;
   - C_F \;\frac{2}{D_1\,D_2\,D_3\,D_7}\,\left(  \,(1-\epsilon)\,\bar{ \alpha }_1\,
          \bar{ \alpha }_2\,\bm{A}_1 \,\bm{B}_2   + 
       \bar{ \alpha }_1\,\bar{ \alpha }_2\,\left( \bm{k} + \bm{r} \right) \,
        \left( \bm{k} + \bm{\ell} + \bm{r} \right)  \right)  \\
\A\B_{13}=&{}\;
  \frac{1}{N_c}\frac{1} {D_1\,D_2\,D_4\,D_6} \left(\,(1-\epsilon)\,\alpha_1^2\, \bm{A}_1\,\bm{B}_3 - \alpha_1^2\,\alpha_2\,\bar{ \alpha }_1\,\bm{r}^2 \right)\\
\A\B_{14}=&{}\;
 N_c \;\frac{1}{D_1\,D_2\,D_5\,D_7}\,\bigg( \frac{1}{2\,
          \alpha_\ell}\,\bar{ \alpha }_1\,
          \left( \alpha_1\,\bar{ \alpha }_1 -\alpha_\ell - (1-\epsilon)\,2\,\alpha_\ell^2 \right) \bm{A}_1\,\bm{B}_4 \nonumber \\*
     &+ \alpha_1\,\bar{ \alpha }_1^2\,
        \bar{ \alpha }_2\,D_2 - 
       \frac{1}{2}\alpha_1\,\alpha_2\,\bar{ \alpha }_1^2\,
          D_7 + 
       \frac{1}{2\,\alpha_\ell}\,\alpha_1\,\alpha_2\,
          \bar{ \alpha }_1\,\bar{ \alpha }_2\, \bm{A}_1 \bm{r}- 
       \alpha_1\,\alpha_2\,\bar{ \alpha }_1\,\bar{ \alpha }_2\,
        \bm{r}^2 \nonumber \\*
      &- \frac{3}{2}\,\bar{ \alpha }_1\,\bar{ \alpha }_2\,
          \bm{A}_1 \left( \bm{k} + \bm{r} \right)\,\bigg) \\
\A\B_{15}=&{}\;
   - N_c \;\frac{1}{\alpha_\ell\,D_1\,D_2\,D_5\,D_7}\alpha_1\,\alpha_2\,\bar{ \alpha }_1^2\,
      \bar{ \alpha }_2\,\bm{r}^2  \\
\A\B_{22}=&{}\;
 \frac{1}{N_c}\;\frac{1}{D_2\,D_3\,D_4\,D_7}\left((1-\epsilon)\,\alpha_2\,\bar{ \alpha }_2\, \bm{A}_2\,\bm{B}_2 + 
     \alpha_2\,\bar{ \alpha }_2\,\left( \bm{k} + \bm{r} - \alpha_1\,\bm{r} \right) \,
      \left( \bm{k} + \bm{\ell} + \bar{ \alpha }_1 \,\bm{r} \right) \right) \\
\A\B_{23}=&{}\;
   - C_F \;\frac{2}{\bar{ \alpha
  }_1\,D_2\,D_4^2\,D_6}\,\,(1-\epsilon)\,\alpha_1^2\,\alpha_2\,\bm{A}_2 \bm{B}_3\\
\A\B_{24}=&{}\;
   - N_c\;\frac{1}{D_2\,D_4\,D_5\,D_7}\,\bigg( \frac{1}{2\,\alpha_\ell} \,\alpha_2\,
            \left( \alpha_1\,\bar{ \alpha }_1 -\alpha_\ell - 2\,(1-\epsilon)\,\alpha_\ell^2 \right) \, \bm{A}_2 \,\bm{B}_4 \nonumber\\*  
  & + \alpha_1\,\alpha_2\,\bar{ \alpha }_1\,\bar{ \alpha }_2\,
        D_2 - \frac{1}{2} \alpha_1\,\alpha_2^2\,
          \bar{ \alpha }_1\,D_7 + 
       \frac{1}{2\,\alpha_\ell} \,\alpha_1\,\alpha_2\,
          \bar{ \alpha }_1\,\bar{ \alpha }_2\, \bm{A}_2 \bm{r} + 
       \alpha_1\,\alpha_2\,\bar{ \alpha }_1\,\bar{ \alpha }_2\,
        \bm{r}^2 \nonumber \\* 
   &- \frac{3}{2} \,\alpha_2\,\bar{ \alpha }_2\,
          \bm{A}_2 \left( \bm{k} + \bm{r} \right)\,\bigg) \\
\A\B_{25}=&{}\;
 N_c\;\frac{1}{\alpha_\ell\,D_2\,D_4\,D_5\,D_7}\alpha_1\,\alpha_2^2\,\bar{ \alpha }_1\,\bar{ \alpha }_2\,\bm{r}^2\\
\A\B_{33}=&{}\;
 \frac{1}{N_c}\; \frac{1}{D_1\,D_3\,D_4\,D_6}\left((1-\epsilon)\,\alpha_1\,\bar{ \alpha }_1\, \bm{A}_3\,\bm{B}_3 + 
     \alpha_1\,\bar{ \alpha }_1\,\left( \bm{k} + \bm{r} - \alpha_1\,\bm{r} \right) \,
      \left( \bm{k} + \bm{\ell} + \bar{ \alpha }_1 \,\bm{r} \right) \right) \\
\A\B_{34}=&{}\;
   - N_c \,\frac{1} {D_1\,D_3\,D_5\,D_7}\,
       \bigg( \frac{1}{2\,\alpha_1\,\alpha_\ell}\bar{ \alpha }_1^2\,
            \left( \alpha_1\,\bar{ \alpha }_2 + 2\,(1-\epsilon)\,\alpha_\ell^2  
              \right) \,\bm{A}_3\,
            \bm{B}_4 \nonumber \\*
     &+ \frac{1}{2}\alpha_1\,\bar{ \alpha }_1^2\,\bar{ \alpha }_2\,D_7 + 
         \frac{1}{2\,\alpha_\ell}\alpha_1\,\bar{ \alpha }_1^2\,\bar{
           \alpha }_2\,\bm{A}_3 \bm{r} + 
         \alpha_1\,\bar{ \alpha }_1^2\,\bar{ \alpha }_2\,\bm{r}^2 \bigg)\\
\A\B_{35}=&{}\;
 N_c \;\frac{1}{\alpha_\ell\,D_1\,D_3\,D_5\,D_7}\alpha_1\,\bar{ \alpha }_1^2\,\bar{ \alpha }_2^2\,\bm{r}^2\\
\A\B_{44}=&{}\;
   - N_c \,\frac{1}{D_2\,D_5^2\,D_7} \,\bigg( \left(  2\,\alpha_1\,\bar{ \alpha }_1 + \alpha_\ell - 2\,(1-\epsilon)\,\alpha_\ell^2\right)  \bm{A}_4\,\bm{B}_4 \nonumber \\*
     &+ \frac{1}{2} \alpha_1\,\alpha_\ell\,\bar{ \alpha }_1\,
          \bar{ \alpha }_2\,D_2 + 
       \frac{1}{2} \alpha_1\,\alpha_2\,\alpha_\ell\,
          \bar{ \alpha }_1\,D_7 +  
    2\,\alpha_1\,\alpha_2\,\bar{ \alpha }_1\,\bar{ \alpha }_2\,
       \bm{r}^2 - \alpha_\ell^2\,\left( \bm{k} + \bm{\ell} \right)
   \,\left( \bm{k} + \bm{r} \right) 
       \bigg) \\
\A\B_{45}=&{}\;
  N_c\;\frac{1}{D_2\,D_5^2\,D_7}\alpha_1\,\alpha_2\,\bar{ \alpha
   }_1\,\bar{ \alpha }_2\,\bm{r}^2  \\
\A\B_{55}=&{}\;0 \;.
\label{eq:AB55}
\end{align}
Here we have used the following abbreviations:
\begin{align}
\bm{A}_1=&\,\alpha_2\, \bm{\ell} + \alpha_\ell \, (\bm{k}+\bm{r})\\
\bm{A}_2=&\,\bar{\alpha}_1\, \bm{\ell} + \alpha_\ell \, (\bm{k}+\bm{\ell})\\
\bm{A}_3=&\,\alpha_1\, \bm{\ell} - \alpha_\ell \, (\bm{k}+\bm{r})\\ 
\bm{A}_4=&\,\alpha_2\, (\bm{\ell} - \bm r) + \alpha_\ell \, (\bm{k}+\bm{r})\\
\bm{B}_1=&\,\bar{\alpha}_2\, \bm{\ell} - \alpha_\ell \,
 (\bm{k}+\bm{\ell})\\
\bm{B}_2=&\,\alpha_1\, \bm{\ell} - \alpha_\ell \, (\bm{k}+\bm{r})\\
\bm{B}_3=&\,\bar{\alpha}_1\, \bm{\ell} + \alpha_\ell \,
 (\bm{k}+\bm{\ell})\\
\bm{B}_4=&\,\bar{\alpha}_2\, (\bm{\ell} - \bm r) - \alpha_\ell \,
(\bm{k}+\bm{\ell})\;.
\label{eq:b4def}
\end{align}
In the above definitions, the vectors $\bm B_i$ are obtained from the
$\bm A_i$ by the transformations \eqref{eq:barr1}, \eqref{eq:barr2}.
Note also the additional relations 
\begin{equation}
  \label{eq:a2b3}
  \bm A_2 = \bm B_3\;, \qquad \bm A_3 = \bm B_2\;.
\end{equation}

The remaining terms $\A\A_{ij}$ (with $i>j$) and $\B\A_{ij}$ (with
$i>j$) follow from the symmetry properties $\A\A_{ij}=\A\A_{ji}$ and
$\A\B_{ij}=\B\A_{ji}$; in order to find $\B\A_{ij}$ and $\A\B_{ji}$
(with $i<j$), and all the $\B\B_{ij}$'s we use the substitutions
\eqref{eq:barr1}, \eqref{eq:barr2}.

We finally note that in the sum of the four $(ij)=(44)$ matrix elements:
$\A\A_{44}+\A\B_{44}+\B\A_{44}+\B\B_{44}$, the term proportional to
$D_2$ in $\A\A_{44}$, the terms proportional to $D_2$ and $D_7$ in
$\A\B_{44}$, and the analogous terms in $\B\A_{44}$ and $\B\B_{44}$
cancel against each other.

An important feature of the results \eqref{eq:AA11}--\eqref{eq:AB55}
is the ultraviolet behavior of transverse momenta: individual terms,
namely $\A \A_{11}$ and $\A \A_{22}$, $\A \A_{12}$ and $\A \A_{21}$,
and $\A \A_{14}$, $\A \A_{24}$, $\A \A_{41}$, $\A \A_{42}$, and $\A
\A_{44}$ go as $1 / \bm \ell^2$ for $\bm \ell^2 \gg \bm k^2$. When
integrating over $\bm \ell$, they lead to logarithmic divergencies.
However, in the sum these divergencies cancel, and all transverse
momentum integrals are ultraviolet finite.

%%%%

\section{The central region and the energy scale}

\noindent
As we have discussed in the beginning of section II, our results are valid 
in the fragmentation region of the photon and in the `upper' part of the 
central region.
Before we move on to any further evaluation of our results we have to address  
the question of how to remove the central region.   
Let us, therefore, first study the limit where the emitted gluon
belongs to the central rapidity region:
\begin{align}
  \label{eq:central}
  \frac{|\bm\ell|}{\sqrt{s}} \le \alpha_{\ell} \le \alpha_{\ell\; 0}
   \nonumber \\ 
  \frac{\bm\ell^2}{s \alpha_{\ell\; 0}}\le \beta_\ell \le 
\frac{|\bm\ell|}{\sqrt{s}}\;.
\end{align}
In this limit we should find agreement with the leading logarithmic
approximation.  First of all, we note that the phase space integral in
\eqref{eq:ps3} contains a factor $1/\alpha_{\ell}$: this makes the
$\alpha_{\ell}$ integral divergent, since, as we will see, some of the
$\A\A_{ij}$ have a finite but non vanishing limit for $\alpha_{\ell}\to
0$. Let us see how this works in detail.  In the limit of small
$\alpha_{\ell}$, the denominators $D_1,\dots ,D_7$
(\eqref{eq:D1}--\eqref{eq:D7}) are expressed in terms of the two
denominators $D_2$ and $D_7$; they in turn are closely related to the
denominators $D(\bm k + \bm r)=\alpha (1-\alpha)Q^2 + (\bm k + \bm
r)^2$ and $D(\bm k + \bm\ell)=\alpha (1-\alpha)Q^2 + (\bm k +
\bm\ell)^2$, that are well known from the Born approximation:
\begin{align}
  \label{eq:dinrholimit}
  D_1 &\to - \frac{1-\alpha}{\alpha_\ell} \bm\ell^2\nonumber\;,\\ 
  D_2 &\to - \frac{1}{\alpha} D(\bm k+\bm r)\nonumber\;,\\ 
  D_3 &\to  \frac{\alpha}{\alpha_\ell} \bm\ell^2\nonumber\;,\\ 
  D_4 &\to \frac{1-\alpha}{\alpha_\ell} \bm\ell^2\nonumber\;,\\ 
  D_5 &\to - (\bm\ell - \bm r)^2 \nonumber\;,\\ 
  D_6 &\to - \frac{\alpha}{\alpha_\ell}  \bm\ell^2\nonumber\;,\\ 
  D_7 &\to - \frac{1}{1-\alpha} D(\bm k + \bm\ell)\;. 
\end{align}
In the numerators on the rhs of \eqref{eq:AA11}--\eqref{eq:AB55} 
we expand in powers of
$\alpha_\ell$ and keep only the leading term. The
contributions from the diagrams containing $A_3$ or $B_3$ drop out
completely, since they involve an extra fermion propagator that gives
a suppression by a power of $\alpha_\ell$. Similarly the diagrams with
$\A_1$, $\A_2$ (or $\B_1$, $\B_2$) on both sides. Nonzero
contributions come from diagrams involving $\A_4$ or $\B_4$, but they
cancel in the sum.  We are finally left with the contributions coming
from $\A_5$ and $\B_5$.  As to the $\A\A_{ij}$-terms, all
non vanishing terms have the common factor $\frac{1}{(D_2)^2}$; all
non vanishing $\A\B_{ij}$ terms contain $\frac{1}{D_2 D_7}$. For the
sum on the rhs of \eqref{eq:m2} we find the simplified form
\begin{equation}
  \label{eq:limitm2}
  |{\cal M}|^2 =
  \frac{2s^2}{\bm{r}^4}\, \frac{2(2\pi)^3}{3} g^6 \frac{\delta^{ab}}{2N_c} N_c\,\delta^{ab}\;
  \frac{\alpha (1-\alpha) |\Psi_L(\bm k + \bm\ell, \alpha)|^2}{(\bm\ell -\bm r)^2} 
  \sum_{i,j=1}^{5} {\cal D}_{ij}\;.
\end{equation}
Here, 
\begin{equation}
|\Psi_L(\bm k+ \bm\ell, \alpha)|^2 = \frac{6
  \alpha_{\mathrm{em}}e_f^2}{\pi^2} \alpha^2 (1-\alpha)^2 Q^2
\left( \frac{1}{D(\bm k + \bm r)} - \frac{1}{D(\bm k + \bm\ell)} \right)^2
\end{equation}
is the square of the leading order (longitudinal) photon wave
function.  The matrix ${\cal D}_{ij}$ does not discriminate between
the coupling of the $t$-channel gluon to the upper quark or antiquark
lines, i.e.\ the difference between $\A_i$ and $\B_i$ is contained
entirely inside the photon wave function. The matrix itself has the
rather simple form
\begin{equation}
  \label{eq:limitmatrix}
  \begin{pmatrix}
    0 & 0 & 0 & -\frac{1}{2} & \frac{\bm r^2}{\bm\ell^2}\\
    0 & 0 & 0 & -\frac{1}{2} & \frac{\bm r^2}{\bm\ell^2}\\
    0 & 0 & 0 & 0 & 0 \\
    -\frac{1}{2} & -\frac{1}{2} & 0 
      & {2+2 \frac{\bm r^2}{( \bm\ell - \bm r )^2}} &
      -\frac{\bm r^2}{( \bm\ell - \bm r )^2}\\ 
    \frac{\bm r^2}{\bm\ell^2}& \frac{\bm r^2}{\bm\ell^2} & 0 & -\frac{\bm r^2}
    {( \bm\ell - \bm r )^2} & 0
\end{pmatrix}\;.
\end{equation}
The sum of all diagrams is then given by
\begin{equation}
  \label{eq:simplm2}
  |{\cal M}|^2 =
  2s^2 \frac{2(2\pi)^3}{3} g^4\delta^{ab}\,
  \alpha (1-\alpha)\,|\Psi_L(\bm k + \bm\ell, \alpha)|^2\,
  \frac{1}{(\bm\ell -\bm r)^4} \,
  \frac{4N_c\,\bm{r}^2 (\bm\ell -\bm r)^2}{\bm\ell^2}\,
  \frac{1}{\bm r^4}\,
  \frac{g^2\delta^{ab}}{2N_c}\;.
\end{equation}
This is exactly what we expect: the factor $\alpha (1-\alpha)
|\Psi_L|^2$, including the color factor $\delta^{ab}$ gives the impact
factor for virtual photons in the LLA. The factors $1/(\bm r
-\bm\ell)^4$ and $1/\bm r^4$ are the propagators of the reggeized
gluons in the $t$-channel, and $4N_c \bm r^2 (\bm
r-\bm\ell)^2/\bm\ell^2$ is the square of the Lipatov vertex
\cite{centralregion}. Finally, $\delta^{ab}/(2N_c)$ is the LLA quark impact
factor (averaged (summed) over initial (final) colors and helicities).
Together with the $\alpha_{\ell}$ integral in the region
\eqref{eq:central} which gives a ln-$s$ factor, we recover the LO BFKL
result.  This demonstrates that our results correctly reproduce the
central region of the produced gluon.  It may be interesting to note
that \eqref{eq:simplm2} may also be written in the factorized form:
\begin{align}
  \label{eq:simplm2fact}
  |{\cal M}|^2 =&
  2s^2 \frac{2(2\pi)^3}{3} g^4\delta^{ab}\,
  \alpha (1-\alpha)\,|\Psi_L(\bm k + \bm\ell, \alpha)|^2\,
  \nonumber\\
  &\times 
  4N_c \left[
    \frac{\bm r- \bm\ell}{(\bm r- \bm\ell)^2}
    +     \frac{\bm\ell}{\bm\ell^2}\,
  \right]\cdot
  \left[
    \frac{\bm r- \bm\ell}{(\bm r- \bm\ell)^2}
    +     \frac{\bm\ell}{\bm\ell^2}\,
  \right]
  \frac{1}{\bm r^4}\,
  \frac{g^2\delta^{ab}}{2N_c}\;.
\end{align}
In this way it has the same structure as
\eqref{eq:AA11}--\eqref{eq:AB55}: a product of two terms, one
belonging to the incoming photon, one to the outgoing. 

In order to avoid double counting we have to remove this central region 
from our results. Details of this subtraction procedure will be presented
elsewhere, here we only sketch the general argument
\cite{CCS,DimitriPhD}. 
Consider, as an example,
the contribution $\A\A_{15}$ which, as $\alpha_{\ell}\to 0$, has a non vanishing
limit. The integral over $\alpha_{\ell}$ diverges logarithmically, but the 
$\ln s$-piece belongs to the LO BFKL calculation. We therefore decompose:
\begin{align}
\label{eq:subtr}
\int_{\frac{|\bm l|}{\sqrt{s}}}^{1-\alpha}
\frac{d\alpha_{\ell}}{\alpha_{\ell}} \A\A_{15}(\alpha_{\ell}) =&
\\
\int_{\frac{|\bm l|}{\sqrt{s}}}^{1-\alpha}
\frac{d\alpha_{\ell}}{\alpha_{\ell}}& \left[ \A\A_{15}(\alpha_{\ell})
  - \A\A_{15}(0) \theta(\alpha_{\ell\;0}- \alpha_{\ell})\right] +
\A\A_{15}(0) \ln(\sqrt{s} \frac{\alpha_{\ell\;0}}{|\bm \ell|})\;,
\nonumber
\end{align}
where $\alpha_{\ell\;0}$ denotes an energy independent scale much
smaller than unity which may depend upon the other momenta, e.g. $\bm
\ell$. As an example, we could define, as the boundary between the
fragmentation region of the photon and the central region, a cutoff on
the invariant mass of the $q \bar q g$ system, $M_0^2$, which we
choose to be larger than $Q^2$, $\bm \ell^2$, $\bm k^2$ and $\bm r^2$. In
this case $\alpha_{\ell\;0} \approx \frac{\bm \ell^2} {M_0^2}$.  In the
first term on the rhs, we can safely put to zero the lower limit of
integration (the error is of order $1/\sqrt{s}$), the second term is
part of the LO BFKL calculation, now with a definite scale inside the
logarithm.  This example illustrates the r\^ole of the energy scale
within the result of the NLO calculation: a change in the scale
results in a finite term. As an example, starting from
\eqref{eq:subtr} and adding the analogous contribution from the
`lower' part of the central region (where we have an analogous cutoff
parameter $\beta_{\ell\;0}$), we can write:
\begin{eqnarray}
\ln(\sqrt{s} \frac{\alpha_{\ell\;0}}{|\bm \ell|})+
\ln(\sqrt{s} \frac{\beta_{\ell\;0}}{|\bm \ell|}) =
\ln(\frac{s}{|\bm r||\bm r - \bm \ell|}) + 
\ln(\frac{|\bm r||\bm r - \bm \ell|\alpha_{\ell\;0} \beta_{\ell\;0}}{\bm \ell^2}).
\end{eqnarray}      
From this we see that our NLO correction does not fix the energy
scale, but, for a given choice of the scale, it provides a
well-defined answer.  Moreover, it then determines how this answer
changes under a modification of the energy scale.

This feature becomes more transparent if we generalize to infinite order  
in $\alpha_s$. Let us consider, instead of 
the process $\gamma^* q \to (q \bar q g)q$, the scattering of two
virtual photons with virtualities $Q_1^2$ and $Q_2^2$, i.e.\ we replace
the lower quark line by another photon impact factor. In the LO
approximation, the scale of the energy factor in the total cross section  
is not fixed, i.e.\ any change of the scale exceeds the LO accuracy. 
One possible choice is the symmetric scale 
$s_0 = kk'$, $k = |\bm k|$, $k' = |\bm k'|$:
\begin{align}
  \label{eq:sigmagsxmpl}
  \sigma_{\gamma^*\gamma^*}(s) =& \frac{1}{(2\pi)^{2}}
    \int_{\delta-i\infty}^{\delta+i\infty}\!
    \frac{d\omega}{2\pi i} 
    \int\! \frac{d^{2}\bm k}{\bm k^2}
    \int\! \frac{d^{2}\bm k'}{{\bm k'}^2} \nonumber\\
    &\times
    \left(\frac{s}{kk'}\right)^{\omega}
    \Phi_{\gamma^*} (\bm k, Q_1)
    \G_\omega (\bm k, \bm k')
    \Phi_{\gamma^*}(\bm k', Q_2). 
\end{align}
This choice seems to be most natural from the point of view of the
Green's function. On the other hand, for a high energy scattering
amplitude it is more natural to chose the `physical' scale
$s_0=\sqrt{Q_1^2 Q_2^2 }$.  In NLO, a change of the scale affects both
the impact factors and the Green's function, i.e.\ the definition of
these quantities becomes scale dependent. For example, starting from
\eqref{eq:sigmagsxmpl} where all elements are defined with respect to
the scale $(kk')$, and changing to the `physical' scale $\sqrt{Q_1^2
  Q_2^2}$ we obtain:
\begin{align}
  \label{eq:sigmagsgs}
  \sigma_{\gamma^*\gamma^*}(s) =& \frac{1}{(2\pi)^{2}}
    \int_{\delta-i\infty}^{\delta+i\infty}\!
    \frac{d\omega}{2\pi i} 
    \int\! \frac{d^{2}\bm k}{\bm k^2}
    \int\! \frac{d^{2}\bm k'}{{\bm k'}^2} \nonumber\\[3pt]
    &\times
    \left(\frac{s}{Q_1Q_2}\right)^{\omega}
    \Phi_{\gamma^*} (\bm k, Q_1)
    \left(\frac{Q_1}{k}\right)^{\omega}
    \G_\omega (\bm k, \bm k')
    \Phi_{\gamma^*}(\bm k', Q_2) 
    \left(\frac{Q_2}{k'}\right)^{\omega}\;.
\end{align}
It seems natural to absorb the additional factors
$(\frac{Q_1}{k})^{\omega}$ and $(\frac{Q_2}{k'})^{\omega}$ into the
impact factors of the photon.  We introduce the Mellin transforms
\begin{align}
  \label{mellinif}
  \Phi_{\gamma^*} (\bm k, Q^2) &=
  \int\!\frac{d\gamma}{2\pi i}
  \left(\frac{Q^2}{k^2}\right)^\gamma \psi(\gamma)\\
  \G_\omega (\bm k, \bm k') &= \int\!\frac{d\gamma}{2\pi i}
  \frac{1}{kk'} \left(\frac{k^2}{{k'}^2}\right)^\gamma \frac{1}{\omega -
    \chi(\gamma)}\;,
\end{align}
where $\chi(\gamma)$ is the eigenvalue function of the BFKL kernel
(for simplicity we ignore the scale dependence of the strong coupling
$\alpha_s$).  Substituting these Mellin transforms in the cross
section formula \eqref{eq:sigmagsgs} and evaluating the integrals
over $\bm k$ and $\bm k'$ (which lead to $\delta$-functions in the
$\gamma$'s), we obtain the following result for the cross section,
\begin{align}
  \label{eq:sigmagamell}
  \sigma_{\gamma^*\gamma^*}(s) =& \frac{1}{4}
  \frac{1}{\sqrt{Q_1^2 Q_2^2}}
  \int_{\delta-i\infty}^{\delta+i\infty}\!
  \frac{d\omega}{2\pi i} 
  \left(\frac{s}{Q_1\,Q_2}\right)^{\omega}\\
  &\times
  \int\! \frac{d\gamma}{2\pi i}
  \left(\frac{Q_1^2}{Q_2^2}\right)^{\gamma}
  \frac{\psi(\gamma - \frac{\omega}{2} -\frac{1}{2}) \psi(-\gamma -
  \frac{\omega}{2}-\frac{1}{2})}{\omega - \chi(\gamma)}\;.
\nonumber
\end{align}
The change of the energy scale therefore leads to shifts in the
$\gamma$ variables \cite{CCS,DimitriPhD}.

An important consistency is the agreement with the renormalization
group equation. If we consider the DIS limit $Q_1^2 \gg Q_2^2$, our
result has to agree with the NLO result of the DGLAP scheme,
restricted to the neighbourhood of the point $\omega=0$ (here
$n=\omega+1$ is the moment index).  It is easy to see that in the
limit $Q_1^2 \gg Q_2^2$ our formula can be rewritten in the following
way:
\begin{equation}
\sigma_{\gamma^*\gamma^*}(s) =
  \frac{1}{4Q_1^2}
  \int_{\delta-i\infty}^{\delta+i\infty}\!
   \frac{d\omega}{2\pi i}
  \left(\frac{1}{x_B}\right)^{\omega}
  \int\! \frac{d\gamma}{2\pi i}
  \left( \frac{Q_1^2}{Q_2^2} \right)^{\gamma} 
  \frac{\psi(\gamma-\omega-1) \psi(-\gamma)}
  {\omega - \chi(\gamma-\frac{\omega}{2}-\frac{1}{2})}\;.
\end{equation}
This last equation provides a stringent test for the choice of the
cutoff parameter of $\alpha_{\ell}$.  Previously, in \cite{CCS,DimitriPhD} the impact factors for on-shell quarks and gluons have
been considered: here the cutoff parameter has been derived from the
angular ordering condition, together with the requirement that the
collinear singularities are reproduced correctly.

\section{The \lowercase{\boldmath $q \bar q g$}-component of the photon wave function}

\noindent
Let us next address the question how our results fit into the picture
of the photon wave function and the color dipole cross section. We expect 
that, if this picture remains valid in NLO, our results should lead to a new 
Fock component 
of the photon, the $q \bar q g$ component. For a precise definition we first 
have to define what we mean, in NLO, by $q \bar q$ component. As an example, 
a $q \bar q g$ state of the photon where the gluon is either soft or collinear 
with the quark (or antiquark), should be counted as a part of the $q \bar q$ 
component. These contributions contain the infrared singularities 
which cancel against those of the virtual corrections. Consequently, 
the NLO definition of the $q \bar q$ component requires a `resolution scale' 
which, in the $q \bar q g$ state of the photon, separates the $q \bar q$ 
component from the $q \bar q g$ component.    

As a first step, we consider our results \eqref{eq:AA11}--\eqref{eq:AB55},
and ignore both the infrared singularities and the subtraction of the central 
region discussed in the previous section. We will show that our results can be
expressed in terms of a new photon wave function, 
$\psi_{q \bar q g}(Q^2; \bm\rho_1, \bm\rho_2, \alpha, \alpha_{\ell})$, and 
the interaction cross section 
$\sigma_{q \bar g g}(\alpha,\alpha_l, \bm \rho_1, \bm \rho_2)$.
To this end, we Fourier transform into transverse
configuration space. We begin with the terms proportional to
$\bm{A}_i\bm{A}_j$, which are present in all $\A\A_{ij}$ (except for
the elements with $j=5$).  As an illustration, consider the
contribution $\A\A_{11}$ to \eqref{eq:m2}. Obviously we can write
\begin{equation}
\label{eq:aa11fact}
\A\A_{11}= \frac{(\bm A_1)_m}{D_1 D_2} 
\left( 2 \delta_{mn} C_F\alpha_1 \bar \alpha_1 \right) 
\frac{(\bm A_1)_n}{D_1D_2}.
\end{equation}
(where the sum over $m$ and $n$ is included). Defining the Fourier
transform of the last factor:
\begin{equation}
\label{eq:ftaa11}
\iint\! \frac{d^2\bm k d^2 \bm \ell}{(2\pi)^2} e^{i\bm \rho_1 \bm k} 
e^{i \bm \rho_2 \bm l} \frac{(\bm A_1)_n}{D_1 D_2}\;,
\end{equation}
and using \eqref{eq:D1}, \eqref{eq:D2}, and \eqref{eq:AA11}, we note
that the dependence of the integrand $\frac{(\bm A_1)_n}{D_1D_2}$ upon $\bm
r$ is only through the combination $\bm k + \bm r$. We therefore
change the integration variable $\bm k$ into $\bm k'= \bm k + \bm r$
and obtain
\begin{equation}
N \sqrt{\frac{Q^2}{\alpha_2\bar\alpha_1}} \iint\! \frac{d^2\bm k d^2 \bm \ell}{(2\pi)^2} e^{i\bm \rho_1 \bm k} 
e^{i \bm \rho_2 \bm l} \frac{(\bm A_1)_n}{D_1 D_2}
= e^{-i \bm \rho_1 \bm r} 
\psi^{(1)}_{q \bar q g;n}(Q^2;\bm\rho_1, \bm\rho_2, \alpha, \alpha_{\ell}), 
\end{equation}
where 
\begin{equation} 
\label{eq:psiqqg}
\psi^{(1)}_{q \bar q g;n}(Q^2; \bm\rho_1, \bm\rho_2, \alpha, \alpha_{\ell}) =
N \sqrt{\frac{Q^2}{\alpha_2\bar\alpha_1}} 
\iint\! \frac{d^2 \bm k' d^2 \bm \ell}{(2\pi)^2} e^{i\bm \rho_1 \bm k'} 
e^{i \bm \rho_2 \bm \ell} \frac{(\bm A_1)_n}{D_1 D_2}
\end{equation} 
has no dependence upon $\bm r$.  Eq.~\eqref{eq:psiqqg} therefore
defines a new Fock component of the (longitudinal) photon wave
function.  $N$ is the normalization constant which can be determined
only in combination with the NLO $q\bar q$ Fock-component. With a
similar result for the first term in \eqref{eq:aa11fact}, $\A\A_{11}$
takes the form:
\begin{align}
\label{eq:aa11}
\frac{Q^2N^2}{\alpha_2\bar\alpha_1}
\A\A_{11}=& \int \frac{{d^2 \bm \rho_1'} {d^2 \bm \rho_2'}}{(2\pi^2)}
e^{i\bm \rho_1' \bm k} e^{i \bm \rho_2' \bm l}
e^{i \bm \rho_1' \bm r} 
\psi^{(1)}_{q \bar q g;n}(Q^2;\bm\rho_1', \bm\rho_2', \alpha, \alpha_{\ell})^*
\left( 2 \delta_{mn} C_F\alpha_1 \bar \alpha_1 \right) \nonumber\\
&\times \int \frac{{d^2 \bm \rho_1} {d^2 \bm \rho_2}}{(2\pi^2)}
e^{-i\bm \rho_1 \bm k} e^{-i \bm \rho_2 \bm l}
e^{-i \bm \rho_1 \bm r} 
\psi^{(1)}_{q \bar q g;n}(Q^2;\bm\rho_1, \bm\rho_2, \alpha, \alpha_{\ell}).
\end{align}
Integrating over the transverse momenta of the $q \bar q g$ system, we 
obtain $\delta$-functions in the $\bm \rho$-vectors which allow to do the
$\bm \rho_1'$, $\bm \rho_2'$ integrals:
\begin{equation}
\label{eq:aa11ft}
Q^2 N^2\int\! \frac{d\alpha\,d\alpha_{\ell}\,d^2 \bm k\,d^2 \bm \ell\,d^2 \bm
  r}{\alpha_1 \bar \alpha_1 \alpha_{\ell}} \A\A_{11} 
=
\int\! 
d\alpha\,d\alpha_{\ell}
\int\! d^2\bm\rho_1\,d^2\bm\rho_2\, 
\left(\psi^{(1)}_{q \bar q g;m}\right)^*\! \sigma^{(11)}_{q \bar q g;mn} 
\psi^{(1)}_{q \bar q g;n} \;,
\end{equation}
where 
\begin{equation}
\sigma^{(11)}_{q \bar q g;mn}(\alpha,\alpha_l, \bm \rho_1, \bm \rho_2)
= \frac{1}{\alpha_\ell}\int d^2 \bm r e^{i\bm \rho_1 \bm r}  
\left( 2 \delta_{mn} \frac{C_F\alpha_1 \bar \alpha_1}{\bm r^4} \right) 
e^{-i \bm \rho_1 \bm r}\;. 
\end{equation}
The final definition of $\sigma^{(11)}_{q \bar q
  g;mn}(\alpha,\alpha_l, \bm \rho_1, \bm \rho_2)$ requires an overall
constant which cannot be determined at this stage.  The incoming and
the outgoing photons have the same transverse coordinates, i.e.\ the
transverse coordinates of the $q\bar qg$ system remain 'frozen' during
the interaction with the lower quark. We consider this as an important
feature of the real corrections.
 
It is easy to see that a similar analysis applies to the terms
proportional to $\bm A_i \bm A_j$ in all other $\A\A_{ij}$.  In
summary, we transform the transverse momenta in the following way: 
\begin{alignat}{2}
  \label{eq:kperptrafo1}
  \frac{\bm A_1}{D_1 D_2}: &\qquad \bm k'= \bm k +\bm r 
  &\qquad&\bm\ell' = \bm\ell\\
  \frac{\bm A_2}{D_2 D_4}: &\qquad \bm k'= \bm k +\bm r 
  &\qquad&\bm\ell'= \bm\ell -\frac{\alpha_{\ell}}{1-\alpha} \bm r\\
  \frac{\bm A_3}{D_1 D_3}: &\qquad \bm k'= \bm k +\bm r 
  &\qquad&\bm\ell' = \bm\ell\\
  \frac{\bm A_4}{D_2 D_5}: &\qquad \bm k'= \bm k +\bm r 
  &\qquad&\bm\ell'= \bm\ell - \bm r\;.
  \label{eq:kperptrafo4}
\end{alignat}
In this way we eliminate the dependence upon $\bm r$ from the Fourier
transforms and obtain expressions which, after insertion into
\eqref{eq:m2}, can be interpreted as photon wave function
components $\psi^{(i)}_{q \bar q g}$.  The $\bm r$ dependence is
entirely carried by the interaction cross sections $\sigma^{ij}_{q
  \bar q g}$, similar as in the $q\bar q$ case.  We carry out the
transformation of the elements $\bm B_i$ in a similar fashion.  The
shifts in the $\bm k$ and $\bm\ell$ integrations are just transformed
according to \eqref{eq:barr1}, \eqref{eq:barr2}.

As the next class of terms, consider the terms proportional to
$\bm{r}^2$.  Starting with $\A\A_{12}$, we proceed as before, but the
Fourier transformation is done only for $\frac{1}{D_1D_2}$ and
$\frac{1}{D_2D_4}$, respectively: using the same shift of momenta as
before, we again find the dependence upon $\bm{r}$ concentrated in
phase factors, and the remaining integrals define new (scalar)
contributions to the photon wave functions $\psi^{(1)}_{q \bar q g}$
and $\psi^{(2)}_{q \bar q g}$ which do not depend upon $\bm{r}$. The
factor $\bm{r}^2$ goes into the cross section $\sigma^{(12)}_{q\bar
  qg}$.

Next those contributions to $\A\A_{ij}$ which contain $D_2$ or $D_7$
in the numerator: the first appearance is in $\A\A_{14}$, and a quick
survey shows that these terms show up only if $i=4$ or $j=4$.
Therefore, these pieces have to be part of the wave function component
$\psi^{(4)}_{q \bar q g}$: e.g.  $\frac{D_2}{D_2D_5}$ or
$\frac{D_7}{D_2D_5}$. Again, the dependence upon $\bm{r}$ can be
factored out and leads to a phase factor which becomes part of the
interaction cross section.

Finally, we have contributions proportional to $\bm A_i \bm{r}$ which
appear only for $j=4$: following again the same steps as before, we
take the vector $\bm A_i$ as a part of the wave function
$\psi^{(i)}_{q \bar q g}$, and we absorb $\bm{r}$ into the cross
section $\sigma^{(i4)}$.  The contributions proportional to scalar
products such as $(\bm{k}+\bm{r})(\bm{k}+\bm\ell)$ are treated in a
similar way: one of the two vectors belongs to the wave function
$(\psi^{(i)})^*$,the other into $\psi^{(j)}$ (there is some freedom of
how this division is done). In any case, the dependence upon $\bm{r}$
can entirely be absorbed into the interaction cross sections
$\sigma_{q \bar q g}$.

As result, all terms appearing on the rhs of $\A\A_{ij}$, $\A\B_{ij}$
etc.  can be cast either into the form \eqref{eq:aa11ft}, or
\begin{equation}
\int\! d\alpha d\alpha_{\ell}
\int\! d^2 \bm \rho_1 d^2 \bm \rho_2 
\left(\psi^{(i)}_{q \bar q g:m}\right)^* \sigma^{(ij)}_{q \bar q g;m} 
\psi^{(j)}_{q \bar q g} 
\end{equation} 
or 
\begin{equation}
\int\! d\alpha d\alpha_{\ell} 
\int\! d^2 \bm \rho_1 d^2 \bm \rho_2 
\left(\psi^{(i)}_{q \bar q g}\right) \sigma^{(ij)}_{q \bar q g} 
\psi^{(j)}_{q \bar q g} \;.
\end{equation}
In other words, we have both scalar and vector-like contributions to the 
new photon wave function $\psi_{q \bar q g}$. 

At this stage, the matrix of interaction cross sections
$\sigma^{(ij)}$ (here the indices $i$ and $j$ have to be extended in
order to include $\A\A$, $\A\B$, $\B\A$, and $\B\B$, i.e.\ they run
from $1$ to $10$) is not diagonal: so far our choice of the basis of
the wave function components had been dictated by the Feynman diagrams
from which we have started our calculations. Since the matrix is
symmetric, we can always find a more physical basis in which the
interactions are diagonal. In this basis, our cross section can be
written as a sum of squares, each of which supports the wave function
interpretation. In particular, the transverse distances in the
incoming and outgoing wave functions are the same: they remain
`frozen' during the interaction with the quark target.  Note that the
definition of eigenstates of the $q \bar q g$-system is determined by
the interaction cross section: it is the interaction with the quark
which filters out the eigenstates of the $q \bar q g$-system.

As we have indicated before, this analysis is not yet complete. 
To proceed further, we have to combine the results of this paper with the 
virtual corrections obtained in \cite{BGQ}: a suitable definition of the 
NLO $q \bar q$ state requires a subtraction of a part of the phase space 
of the $q \bar q g$ final state. Together with the subtraction of the 
central region, this may very well lead to a `soft melting' of the 
`frozen' transverse distances. This would not be totally unexpected:  
a soft gluon occupies a large region in transverse space, and it 
may take the form of a non-local cloud around the quark or antiquark.  

\section{Conclusions}

In this paper we have computed, for a longitudinally polarized virtual 
photon, the helicity-summed squared matrix element of the process
$\gamma^* +q \to (q \bar q g) +q$. The calculation is valid in the Regge limit 
with a large rapidity gap between the $q \bar q g$-system and the other 
outgoing quark. The results are of interest for several different lines of 
research. In each case, further calculations are needed which will be based 
upon the results of this paper.

First of all, our results present the real corrections to the NLO
corrections of the photon impact factor. In order to complete the NLO
calculations we have to integrate over the final momenta of the $q
\bar q g$-system and to combine with the virtual corrections obtained
in \cite{BGQ}.  In the sum of the real and virtual corrections the
infrared singularities have to cancel; furthermore, we have to
subtract the contribution of the central region.

Next, the results of this paper contain the three-parton final state
of the photon-gluon fusion process $\gamma^* +g \to q \bar q g$ where
the incoming gluon is off-shell.  The complete NLO calculation of this
process also requires, in addition to the tree-level calculation of
the process $\gamma^* +g \to q \bar qg$, to consider the sum with the
virtual corrections to $\gamma^* +g \to q \bar q $ (which is contained
in \cite{BGQ}). From the $q\bar qg$ final states we have to subtract
the central region. With a suitable jet definition, one obtains the
two-jet and the three-jet cross sections in the $k_T$-factorization
scheme.

Finally, our results are part of the NLO corrections to the photon
wave function picture. Ignoring infrared singularities and the problem
of the central region, we have calculated the new $q \bar q g$ Fock
component of the photon wave function. However, for a proper treatment
of the infrared part we need to consider also the virtual corrections
in \cite{BGQ}: we have to define a NLO $q \bar q$ Fock state which
includes a part of the soft gluons in our $q \bar q g$ final state.

\bigskip 

\emph{Note added:} Recently, a paper by V.~S.~Fadin, D.~Y.~Ivanov and
M.~I.~Kotsky has appeared which also addresses the NLO calculation of
the virtual photon impact factor (hep-ph/0106099).  It summarizes the
status of the virtual corrections, repeating the content of
hep-ph/0007119, and it presents the matrix elements for the process
$\gamma^*+q\to (q\bar qg)+q$. In our paper we have calculated the
helicity sum of the squared matrix elements for this process. 

\acknowledgments
\noindent Very helpful discussions with D.\ Colferai, J.\ Collins and 
C.-F.\ Qiao are gratefully acknowledged. One of us (J.B.) has profited
from a useful discussion with M.\ Ciafaloni.

\end{document}